\documentclass[aps,preprint]{revtex4}
\usepackage{amsfonts}
\usepackage{amsmath}
\usepackage{amssymb}
\usepackage{graphicx}

\begin{document}

\title{Interaction of the Torque-Induced Elastic Charge and Elastic Dipole
with a Wall in a Nematic Liquid Crystal. }
\author{V. M. Pergamenshchik and V. A. Uzunova}
\affiliation{$^{1}$Korea University, Display\&Semiconductor Physics, Jochiwon-eup,
Yeongi-gun, Chungnam 339-700, South Korea}
\affiliation{$^{2}$Institute of Physics, prospect Nauki, 46, \ Kiev 03039, Ukraine}
\date{\today}

\begin{abstract}
We show that the elastic charge of colloids in a nematic liquid
crystal can be generated by the vector of external torque. The
torque components play the role of two component charge (dyad) and
give rise to the Coulomb-like potential, while their conservation
law plays the role similar to that of Gauss' theorem in the
electrostatics. The theory is applied to the colloid-surface
interaction. A wall with homeotropic or planar director is shown to
induce a repulsive $1/r^{4}$ force on the elastic dipole. The
external torque, however, induces the elastic charge in this colloid
and triggers switching to the $1/r^{2}$ repulsion.


Short title: elastic charge-wall interaction.
\end{abstract}

\maketitle

\section{Introduction}

Particles of a submicron and micron size immersed in a nematic liquid
crystal (NLC) interact via the director field $\mathbf{n}$ which mediates
the distortions induced by their surfaces \cite%
{Lopatnikov,Terentev0,Kuksenok,Prost,Terentev,Poulin1,Lubensky,LevTomchuk,Poulin2,Stark,Lev2}%
. The new field of nematic colloidal systems, or nematic emulsions \cite%
{Stark}, has gained a continuous growing interest over the past few years.
The physics of these anisotropic colloidal systems has a deep similarity to
the electrostatics. It has been shown that the director-mediated interaction
is of a long range and possesses many other properties characteristic of the
interaction between electric dipoles and quadrupoles \cite%
{Lopatnikov,Terentev0,Kuksenok,Prost,Terentev,Poulin1,Lubensky,LevTomchuk,Poulin2,Stark,Lev2}%
. Particle trapping techniques \cite{MusevicPRL2004} have been used to test
this analogy and demonstrate experimentally the dipole-dipole \cite%
{YadaPRL2004,Smalyukh2004}, quadrupole-quadrupole \cite{PRL2005,CopicPRL2006}%
, and mixed disclination-dipole \cite{Galerne} pair interactions.
Reorientation of the elastic dipoles was shown to be responsible for phase
transitions between different 2-dimensional colloidal lattices on a
nematic-air interface \cite{HL3}. In the context of this analogy\ it is
natural to expect that the Coulomb interaction, which is fundamental to the
electrostatics, has an important implication in the physics of nematic
emulsions, too. Recently\ we developed the electrostatic analogy in nematic
emulsions to the level of charge and its density \cite{Vera1,Vera2}. The
director-mediated Coulomb-like interaction of two colloids was shown to be
fully determined by vectors $\mathbf{\Gamma }_{\bot }^{(1)}$ and $\mathbf{%
\Gamma }_{\bot }^{(2)}$ of the transverse external torques (perpendicular to
the unperturbed director at infinity)\ applied on the colloids \cite%
{Vera1,Vera2}. The scalar product $\ -(\mathbf{\Gamma }_{\bot }^{(1)}\cdot
\mathbf{\Gamma }_{\bot }^{(2)})$ plays the role of the product of two
electrostatic charges in the $1/r$ interaction potential, and thus the two
components of external torque play the role of two component elastic charge.
Because of the difference between the scalar electrostatics and vector
nematostatics, \ the elastic analogues of the surface charge density,
charge, and higher multipole moments consist of two tensors (dyad). The
multipole moments are naturally expressed via the elastic charge density
which is determined by the two transverse director components on the surface
imposing the director deformations. The interaction of the axially-symmetric
sources, considered phenomenologically in \cite{Lubensky}, obtains as
particular case of the interaction of two correspondent multipole dyads.
Small parameter of the theory is the ratio $a/r=($colloid size/distance
between colloids). For small $a/r$ the theory provides all the tools
available in the electrostatics, e.g., for solving different boundary
problems that can occur in the nematostatics of anisotropic emulsions. In
this paper we apply the nematostatics developed in \cite{Vera1,Vera2} to the
interaction between an elastic charge (dyad) and elastic dipole (dyad) with
a wall (surface bounding the NLC) with different director alignments, which
is the elastic counterpart of the well-known electrostatic problem solved by
the method of images. In the next section we briefly introduce the colloidal
nematostatics of Ref.\cite{Vera1,Vera2} and show that the integral form of
the torque balance plays the role similar to that of Gauss' theorem in the
electrostatics. Then the theory is applied to the colloid-surface
interaction. A wall with homeotropic or planar director is shown to induce a
repulsive $1/r^{4}$ force on the elastic dipole. The external torque,
however, induces the elastic charge in this colloid and triggers switching
to the $1/r^{2}$ repulsion. These results suggest that predictions of the
colloidal nematosctatics can be tested by observing behavior of a single
colloid at a sample surface which, in some situations, can be more robust
than dealing with two colloids.

\section{Elastic charge density representation of the colloidal nematostatics%
}

\subsection{Torque balance, Gauss' theorem, and elastic charge in 3
dimensions.}

The fundamental physical quantity of electric charge is purely
phenomenological and must be \textit{postulated} in the theory of elementary
particles. In contrast, the nematostatics of the director field $\mathbf{n}$
allows for \textit{introduction} of two different charges. Electrostatic
potential is a scalar described by the linear Laplace (or Poisson) equation.
It is the linearity that underlies the definition of the electric charge and
its density as the source of electric field. At the same time, $\mathbf{n}$
is a vector field which reduces to a single variable, described by a linear
equation (in the one constant approximation), only in 2 dimensions (2d).
Owing to the linearity, the deformation source can be straightforwardly
established: core of a point defect plays the role of a charge in 2d \cite%
{deGennes,Kleinert,Chaikin,Petty}. The independence of the integral,
expressing the topological invariant, of the integration contour plays the
role anlogous to Gauss' theorem in electrostatics, the invariant itself
plays the role of a conserved charge, and the 2d nematostatics is similar to
the 2d electrostatics with its logarithmic potential: disclinations of the
same signs repel and those of the opposite signs attract each other.

In 3d, however, the analogy between topological defects and charge is
completely lost. In 3d, the field $\mathbf{n}$ is described by highly
nonlinear equations \cite{deGennes} so that point defects, though remain
topological invariants, cannot be linearly connected with the distortions of
$\mathbf{n}$ they induce \cite{Lubensky,Stark}. Here the deformation sourse
is the director distribution around the particle in its close vicinity of a
size $\sim $ $a.$ We refer to such the deformation domain as particle though
the distortion therein can be induced by surface of a real particle, by
topological defects with zero total topological charge \cite{Lubensky,Stark}%
, or by an external field dying out outside the domain area, Fig. 1. \
Consider 3-d director field $\mathbf{n}(\mathbf{r}),$ uniform and parallel
to the $z$-axis at infinity, $\mathbf{n}_{\infty }=(0,0,1)$. At distances $%
r\gg a,$ the small particle-induced perturbation $n_{t}$ of $\mathbf{n}%
_{\infty }$ is transverse, $t=x,y,$ and (in the one-constant approximations
assumed in this paper) has the form

\begin{equation}
n_{t}(\mathbf{r})=\frac{q_{t}}{r}+3\frac{(\mathbf{d}_{t}\cdot \mathbf{r})}{%
r^{3}}+5\frac{(\mathbf{Q}_{t}:\mathbf{r}:\mathbf{r)}}{r^{5}}+...,  \label{nt}
\end{equation}%
It is natural to identify the coefficients with the subscript $t$ in this
expansion with the $t$-th component of elastic charge, elastic dipole, and
elastic quadrupole, respectively. We seek the elastic analog of charge,
following de Gennes' idea outlined in \cite{deGennes}. A transverse external
torque $\mathbf{\Gamma }_{\bot }=(\Gamma _{x},\Gamma _{y},0)$ applied on a
particle in the equilibrium is balanced by another, the elastic torque
distributed over a surface $S$ enclosing the particle, i.e.,

\begin{equation}
\Gamma _{t}=K\int_{S}\varepsilon _{\alpha t\rho }(r_{\rho }\partial _{\beta
}n_{\gamma }\partial _{\alpha }n_{\gamma }+n_{\rho }\partial _{\beta
}n_{\alpha })dS_{\beta },  \label{Gauss}
\end{equation}%
whrere $K$ is the elastic constant, $\varepsilon _{\alpha t\rho }$ is the
absolute antisymmetric tensor, all indices but $t$ run over $1,2,3,$ and
summation over the repeated indices is implied. The integral in the r.h.s.
does not depend on the choice of enclosing surface $S,$ and the equality (%
\ref{Gauss}) reminds one Gauss' theorem with $\Gamma _{t}$ in place of the
electric charge. To further justify this connection one notices that
integral (\ref{Gauss}) over a remote surface $S$ vanishes for any term in
the expansion (\ref{nt}) but the first one. Substituting $n_{t}=q_{t}/r$ in (%
\ref{Gauss}) and integrating over a large sphere gives $\Gamma _{y}=4\pi
Kq_{x},$ $\Gamma _{x}=-4\pi Kq_{y},$ or

\begin{equation}
q_{t}=\frac{[\mathbf{\Gamma }\times \mathbf{n}_{\infty }]_{t}}{4\pi
K}. \label{q=gamma}
\end{equation}

Thus, the tentative conclusion is that, in 3d, the role of Gauss' theorem
and charge is played, respectively, by the balance of external and elastic
torques (conservation of torque) and by the transverse components of the
external torque exerted on the particle. This is fully justified by
calculating the Coulomb-like interaction in the elastic charge density
representation developed in \cite{Vera1,Vera2}.

\begin{figure}[h]
\includegraphics[height=3.5in]{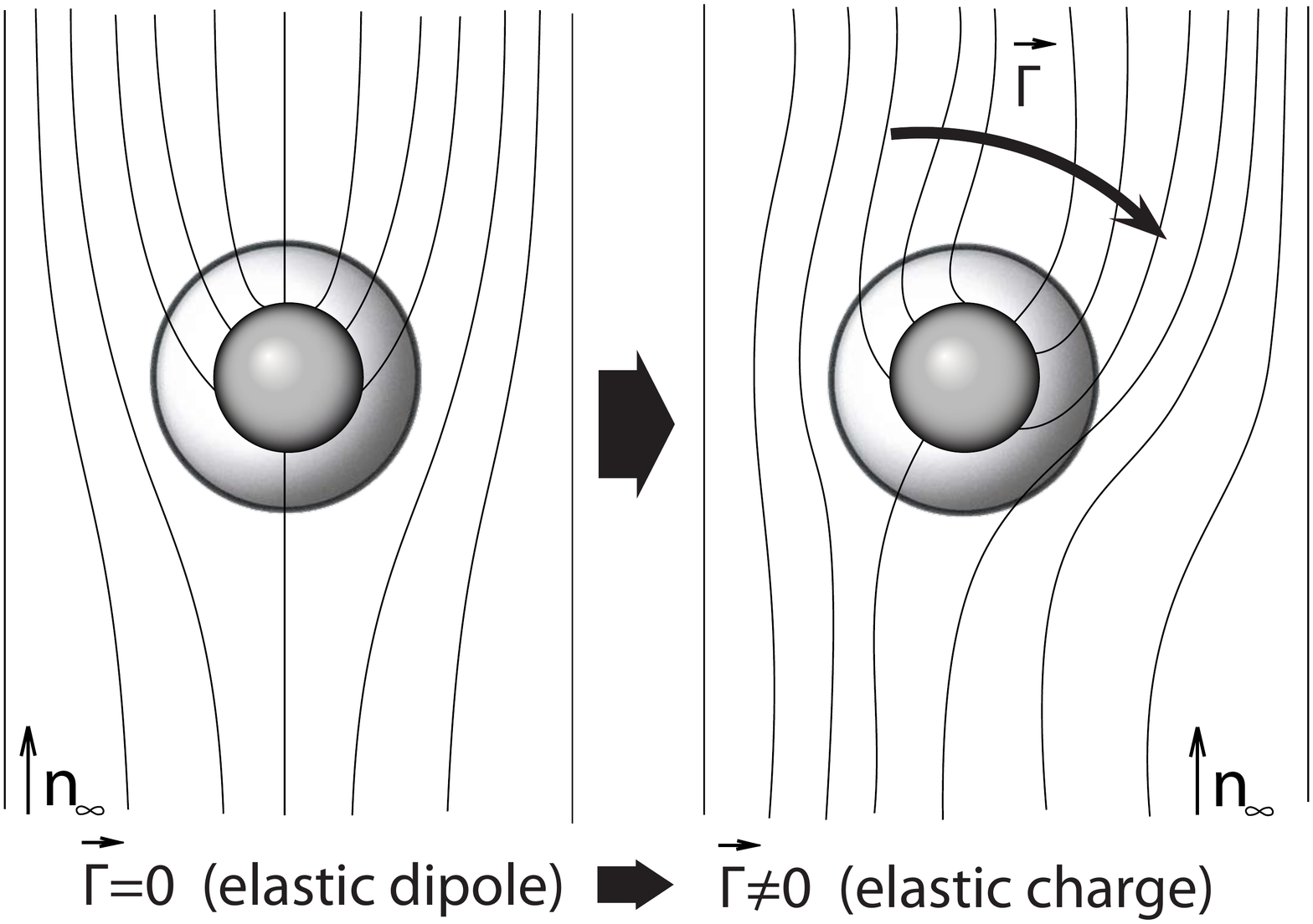}
\caption{General deformation source ("partcile') in 3d. Inside the gray
sphere the deformations can be very large (e.g., induced by point defects or
strong anchoring of a real colloid). But outside the larger sphere the
deformations are weak and linear which allows for the electrostatic anlogy.
The particle itself is of the dipolar type, but an external torque upon it
can charge it, and it becomes an elastic charge.}
\end{figure}

\subsection{Dyads of elastic mutipoles and their interaction via the
director field: the outline}

The results of Refs. \cite{Vera1,Vera2} instructive to our task here can be
summarized as follows. Consider a deformation source with the director
distributions $n_{t}$ given on the surface of enclosing sphere $S$ with
radius $a.$ The quantity%
\begin{equation}
\sigma _{t}(\mathbf{s})=n_{t}(\mathbf{s})/a^{2},  \label{sigma}
\end{equation}%
plays the role of two component surface elastic charge density on the sphere
$S.$ Using natural analogy with the electrostatics, we define the two
component elastic multipoles via the surface charge density as the following
integrals over sphere $S$ enclosing the particle:

\begin{equation}
q_{t}=\frac{a}{4\pi }\int_{S}\sigma _{t}d^{2}s,  \label{q}
\end{equation}

\begin{equation}
d_{t,\alpha }=\frac{a^{2}}{4\pi }\int_{S}\sigma _{t}\nu _{\alpha }d^{2}s,
\label{d}
\end{equation}%
\begin{equation}
Q_{t,a\beta }=\frac{a^{3}}{8\pi }\int_{S}\sigma _{t}(3\nu _{\alpha }\nu
_{\beta }-\delta _{\alpha \beta })d^{2}s,  \label{Q}
\end{equation}%
where $\mathbf{\nu }$ is the vector of a unit outer normal to $S.$ With
these definitions, we obtained the interaction of two particles with similar
multipoles. Interaction of two "charged" particles with nonzero $q_{t}$ is
Coulomb-like:
\begin{equation}
U_{Coulomb}=-4\pi K\frac{q_{t}^{(1)}\cdot q_{t}^{(2)}}{R}=-\frac{(\mathbf{%
\Gamma }_{\bot }^{(1)}\cdot \mathbf{\Gamma }_{\bot }^{(2)})}{4\pi KR},
\label{qq}
\end{equation}%
where we used relation (\ref{q=gamma}), $\mathbf{\Gamma }_{\bot }^{(i)}=%
\mathbf{\Gamma }^{(i)}-(\mathbf{\Gamma }^{(i)}\mathbf{\cdot n}_{\infty })$
is the transverse component of the torques exerted upon the $i$-th
particles, and $R$ is the modulus of the separation vector $\mathbf{R}.$ The
above connection between the elastic charge and external torque is thus
fully justified. Eq. (\ref{qq}) shows that, depending on the sign of $(%
\mathbf{\Gamma }_{\bot }^{(1)}\cdot \mathbf{\Gamma }_{\bot }^{(2)}),$ the
elastic Coulomb interaction can be attractive or repulsive$.$ In contrast to
the electrostatics and 2d nematostatics, the charges with the same sign
attract and with different signs repel each other ("parallel torques"
attract whereas two "antiparallel torques" repel each other). Although the
colloids must be anchored to the director, the Coulomb-like interaction does
not directly depend on their specific shape and anchoring. Instead, the
elastic charge is determined by the coefficients describing the torque
exerted upon the colloid by a given type of external field. For instance,
this can be the vector of permanent electric and magnetic dipole or electric
and magnetic polarizability tensors of a given colloid.

If the external torques are absent, the interaction energy is expressed
solely in terms of particles' multipoles. The interaction between two
"dipolar" particles is of the form

\begin{equation}
U_{dd}=-12\pi K\frac{(\mathbf{d}_{t}^{(1)}\cdot \mathbf{d}_{t}^{(2)})-3(%
\mathbf{d}_{t}^{(1)}\cdot \mathbf{u})(\mathbf{d}_{t}^{(2)}\cdot \mathbf{u})}{%
R^{3}},  \label{Udd}
\end{equation}%
where $\mathbf{u}=\mathbf{R/}R$ is a unit vector along the separation
direction. Eqs.(\ref{nt}),(\ref{sigma})-(\ref{Udd}) (along with the
quadrupole-quadrupole potential derived in \cite{Vera1,Vera2}) suggest the
following interpretation. $q_{t}$ is the $t$-th component of the elastic
charge and $\sigma _{t}(\mathbf{s})$ is its surface density at point $%
\mathbf{s}$ on the sphere. The vector $\mathbf{d}_{t}$ and tensor $\mathbf{Q}%
_{t}$ are the $t$-th dipole and quadrupole moments determined in the
standard way by the surface charge density $\sigma _{t}$ on the sphere$.$ As
$\sigma _{x}$ and $\sigma _{y}$ are separate sources, they determine not
only the $x$ and $y$ director components outside the particle, Eq.(\ref{nt}%
), but also two independent tensors (dyad) for each multipole moment, i.e., $%
q_{x}$ and $q_{y},$ $\mathbf{d}_{x}$ and $\mathbf{d}_{y},\mathbf{Q}_{x}$ and
$\mathbf{Q}_{y},$ and so on.

\section{Colloid-wall interaction in a nematic liquid crystal}

The above formulas can be used to solve boundary problems similar to those
of electrostatics. The simplest boundary problem is the interaction of an
elastic multipole with a surface bounding the nematic sample and imposing
planar or homeotropic director alignment. Here we consider this problem for
an elastic charge and dipole.

\subsection{Repulsion of an elastic charge from the wall}

Let us consider a single particle with a charge at a distance $h$ from a
plane surface of a NLC sample. We assume that the anchoring is strong, the
director alignment in the sample $\mathbf{n}_{\infty }$ far from the
particle is homogeneous and parallel to the $z$-axis, $\mathbf{n}_{\infty
}=(0,0,1),$but the angle it makes to the surfaces is arbitrary. The charge $%
q_{t}$ can be induced by an external field exerting the torque $\mathbf{%
\Gamma }$ with the components $\Gamma _{y}=4\pi Kq_{x}$ and $\Gamma
_{x}=-4\pi Kq_{y}$. To justify the linearized theory, $h$ is assumed to be
large compared to the particle' size. As the director on the sample surface
is fixed, the boundary condition is $\mathbf{n}_{t}=0$, $t=x,y.$

The problem can be solved using the mirror-image method. Let us
place the image-particle with the charge $q_{t}^{^{\prime }}$ on the
other side of a surface at distance $h$ from it, {Figs.2,3.} For
large $h,$ distortions induced by the charge and its image are given
by the sum (see Eq.(\ref{nt})):

\begin{figure}[h]
\includegraphics[height=3in]{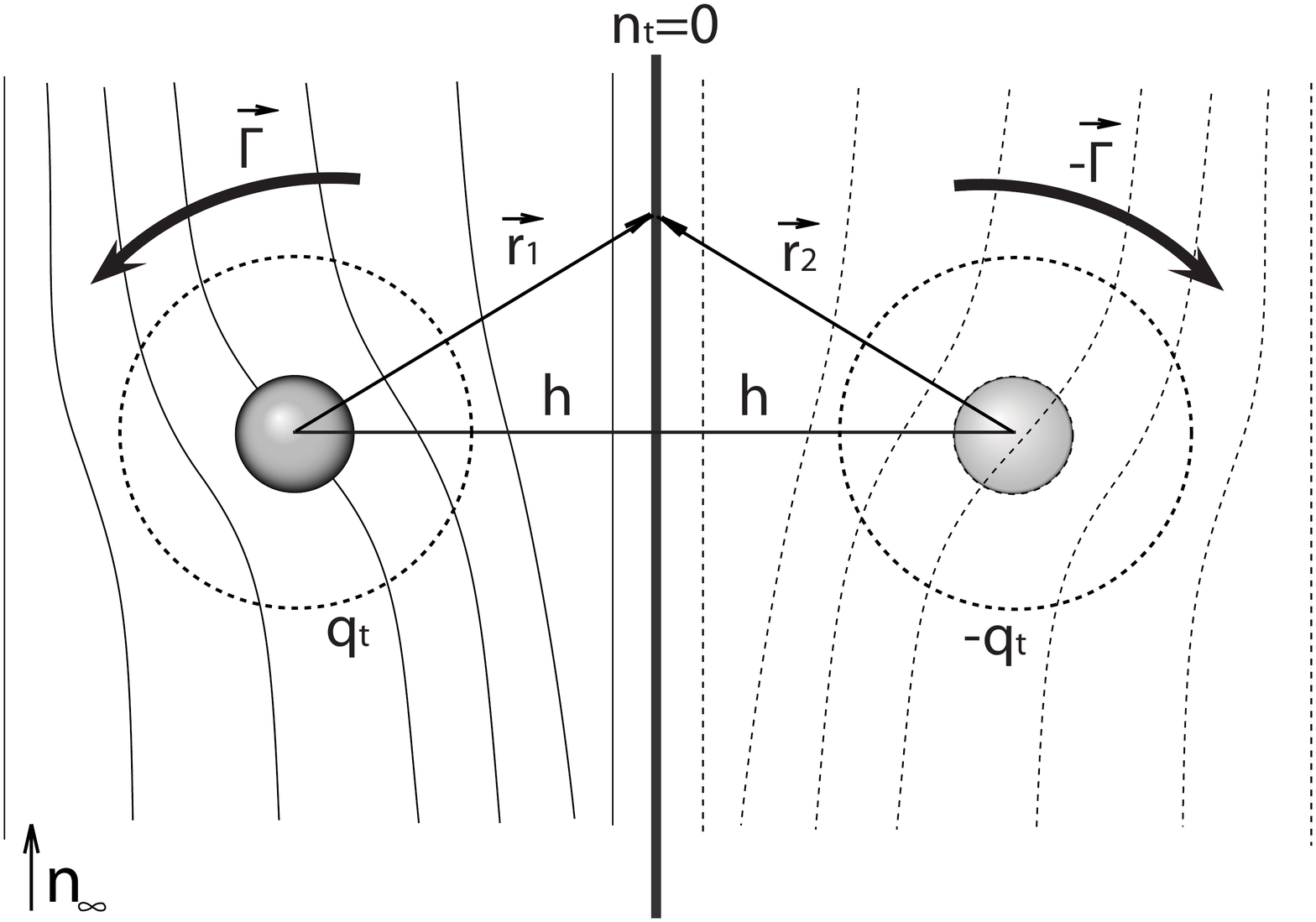} \caption{Elastic charge at a wall with fixed planar director alignment.
Elastic charge $q_{t}$ induced by an external torque $\mathbf{\Gamma
}$ and its image $-q_{t}$ induced by the image-torque
$-\mathbf{\Gamma }$ . The
director at the wall remains unperturbed and equal to $\mathbf{n}_{\infty }.$%
}%
\includegraphics[height=3in]{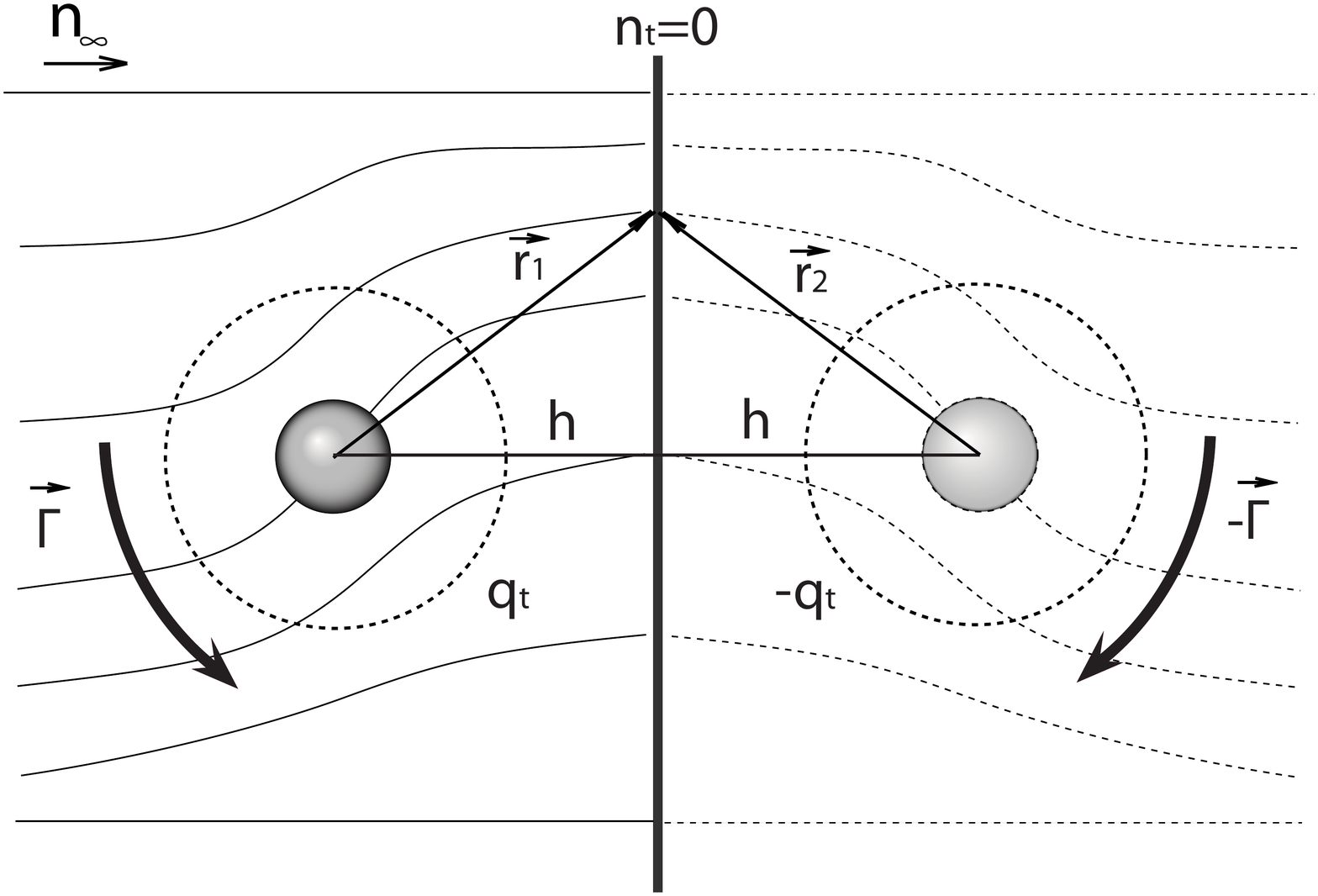}
\caption{Elastic charge at a wall with fixed homeotropic director
alignment. Elastic charge $q_{t}$ induced by an external torque
$\mathbf{\Gamma }$ and its image $-q_{t}$ induced by the
image-torque $-\mathbf{\Gamma }$ . The
director at the wall remains unperturbed and equal to $\mathbf{n}_{\infty }.$%
}
\end{figure}

\begin{equation}
n_{t}(\mathbf{r})=\frac{q_{t}}{r_{1}}+\frac{q_{t}^{^{\prime }}}{r_{2}},
\label{nq}
\end{equation}%
where $r_{1}$ and $r_{2}$ are the distances from a given point of the wall
to the location of the charge and its image. As $r_{1}$ $=r_{2},$ the
boundary condition is satisfied for $q_{t}=-q_{t}^{\prime }$, $t=x,y.$ The
fact that the particle and its image are oppositely charged means that $%
\mathbf{\Gamma }_{\bot }=-\mathbf{\Gamma }_{\bot }^{^{\prime }}$, {%
Figs.2,3.} As two opposite elastic charges repel each other, the elastic
charge-wall interaction is repulsive. The repulsion force obtains from the
interaction energy (\ref{qq}) of the torques $\mathbf{\Gamma }_{\bot }$ and $%
-\mathbf{\Gamma }_{\bot }$ by differentiating with respect to $R$ at $R=2h$,
i.e.,

\begin{equation}
F_{q}=\frac{\mathbf{\Gamma }_{\bot }^{2}}{16\pi Kh^{2}}.  \label{q_force}
\end{equation}

The result depends on the direction of $\mathbf{n}_{\infty }$ and thus on
the surface tilt via the relation $\mathbf{\Gamma }_{\bot }=\mathbf{\Gamma }%
-(\mathbf{\Gamma \cdot n}_{\infty }).$

\subsection{Repulsion of an elastic dipole from the wall}

Now consider the interaction between a wall and an elastic dipole
represented by dyad $(\mathbf{d}_{x},\mathbf{d}_{y}),$ Eq.(\ref{d}). In
general, each of the two vectors $\mathbf{d}_{x}$ and $\mathbf{d}_{y}$ has
three nonzero components. Symmetry makes some of them vanish. For ellipsoids
with one of their axes along $\mathbf{n}_{\infty }=(0,0,1),$ the dipole dyad
is diagonal: $\mathbf{d}_{x}=(d_{x},0,0),\;\mathbf{d}_{y}=(0,d_{y},0)$ where
${d}_{x}\neq {d_{y}}$. In the case of an axially symmetric particle with
symmetry planes passing through the symmetry axis assumed to be along $%
\mathbf{n}_{\infty }$, ${d}_{x}={d_{y}=d}$ (note that an axially
symmetric particle without symmetry planes, such as a helicoid, is a
chiral source which will be considered elsewhere). We restrict our
consideration to this simple and practically important case of
colloids. For instance, such are the so-called "topological
dipoles", i.e., spherical particles with homeotropic boundary
conditions with a companion hyperbolic hedgehog or disclination ring
\cite{Lubensky,Stark}.

From the general equation (\ref{Udd}), the interaction energy of two axially
symmetric dipolar particles with nonzero components $d^{(1)}$ and $d^{(2)}$
obtains in the form

\begin{equation}
U_{dd}=\frac{12\pi Kd^{(1)}d^{(2)}}{R^{3}}(1-3\cos ^{2}{\theta }),
\label{Udd2}
\end{equation}%
where $\theta $ is the angle the separation vector $\mathbf{R}$, which, in
our geometry, is along the surface normal, makes with the far homogeneous
director $\mathbf{n}_{\infty }$. This formula, up to the coefficient $3,$
reproduces the one obtained for the axially symmetric "topological dipoles"
in \cite{Lubensky,Stark}.

\begin{figure}[h]
\includegraphics[height=3.5in]{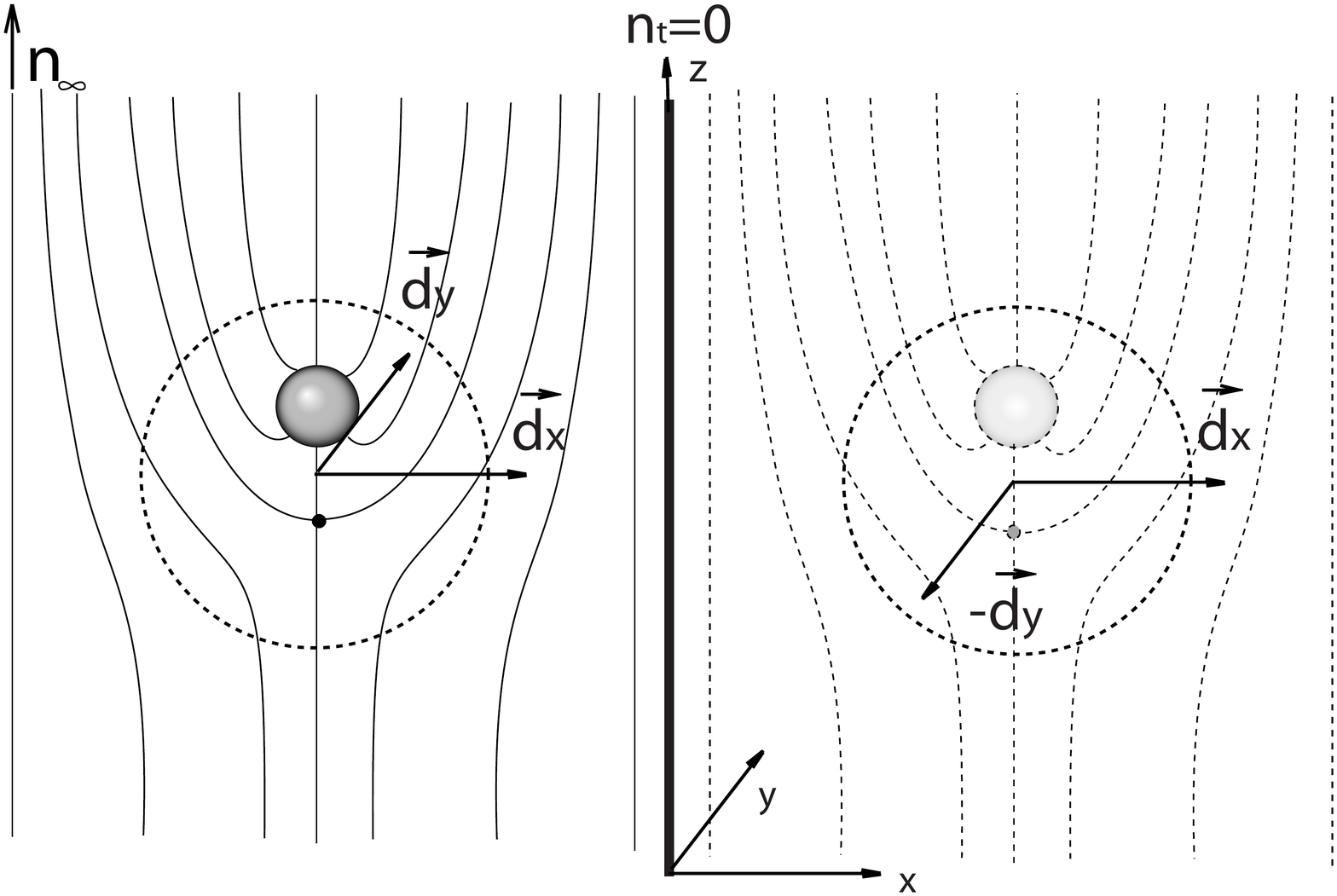}\\
\includegraphics[height=3.5in]{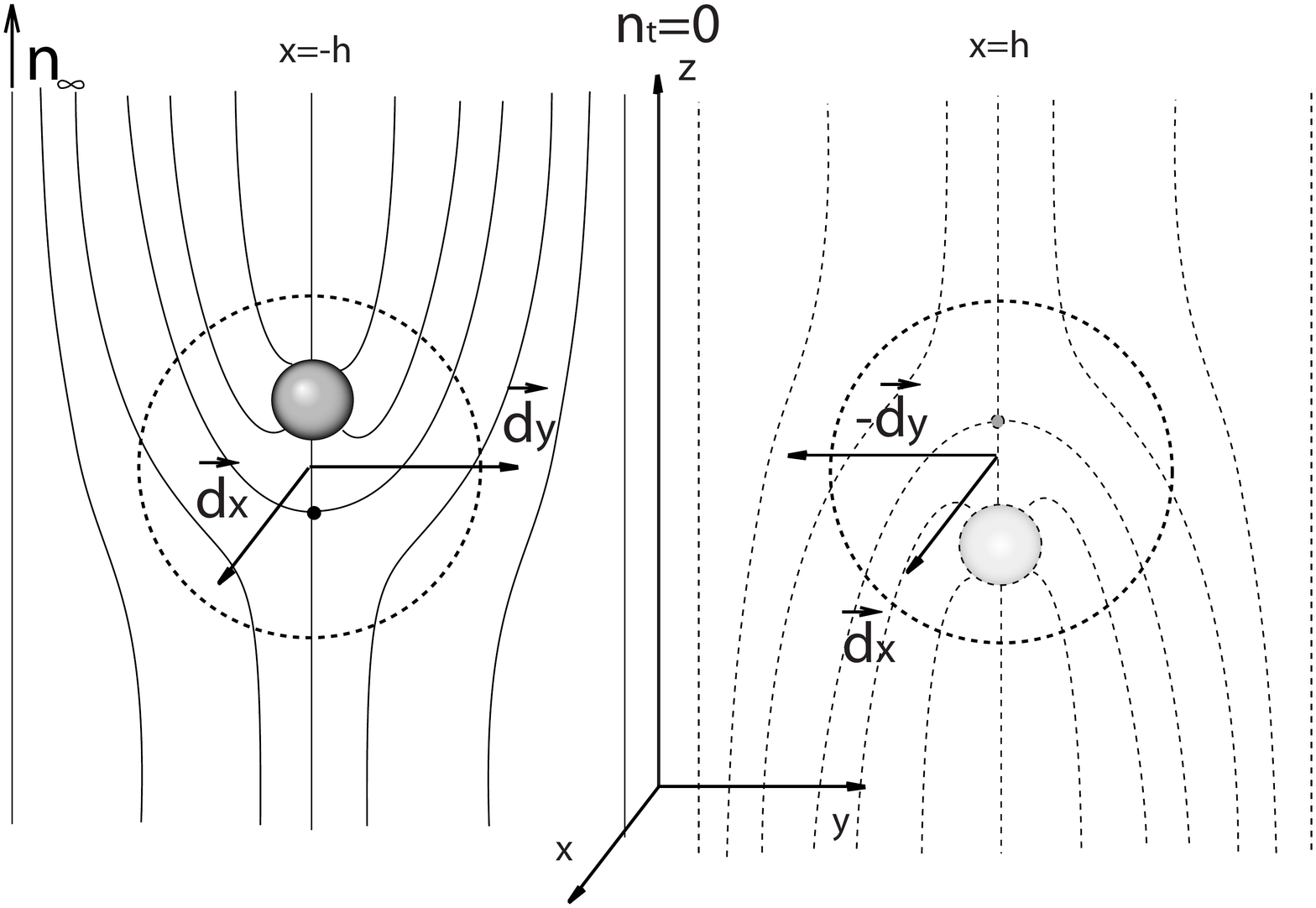}
\caption{Dyad of elastic dipole at a wall with fixed planar director
alignment. The dipole (left) $\ $with $\mathbf{d}_{x}=(d,0,0)$ and$\;\mathbf{%
d}_{y}=(0,d,0)$\ and the image (right) with $\mathbf{d}_{x}^{\prime
}=(d,0,0) $ and$\;\mathbf{d}_{y}^{\prime }=(0,-d,0)$ shown in two
mutually perpendicular planes: a) $xz$-plane normal to the wall and
b) planes $x=-h$ (left) and $x=h$ (right)\ parallel to the wall. }
\label{dd}
\end{figure}

\begin{figure}[h]
\includegraphics[height=3.5in]{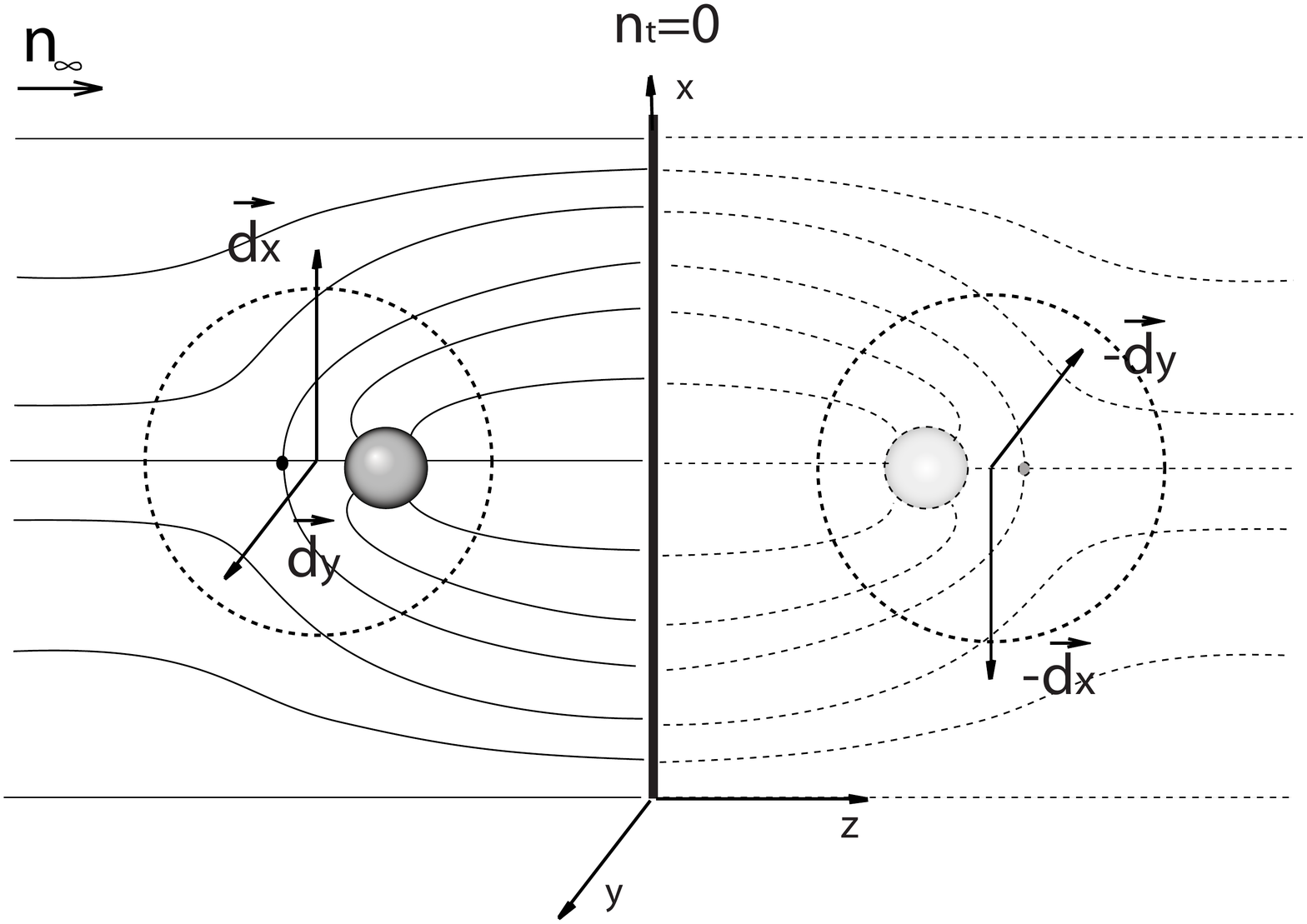}
\caption{ Dyad of elastic dipole at a wall with fixed homeotropic director
alignment. The dipole (left) with $\mathbf{d}_{x}=(d,0,0)$ and$\;\mathbf{d}%
_{y}=(0,d,0)$\ and the image (right) with $\mathbf{d}_{x}^{\prime }=(-d,0,0)$
and$\;\mathbf{d}_{y}^{\prime }=(0,-d,0)$ shown in the $xz$-plane normal to
the wall.}
\end{figure}

Consider an elastic dipole dyad $\mathbf{d}_{t},$ $t=x,y,$ at a distance $h$
from the sample surface (wall). The $z$-direction is along the unperturbed
homogeneous director $\mathbf{n}_{\infty }=(0,0,1).$The general boundary
condition on a surface with strong anchoring of any type is again $\mathbf{n}%
_{t}=0$, $t=x,y.$ The image-dipole $\mathbf{d}_{t}^{^{\prime }}$ is located
at the distance $h$ on the opposite side of the wall. The director field at
point $\mathbf{r}$ of the surface is

\begin{equation}
n_{t}(\mathbf{r})=3\frac{(\mathbf{d}_{t}\cdot \mathbf{r}_{1})}{r_{1}^{3}}+3%
\frac{(\mathbf{d}_{t}^{^{\prime }}\cdot \mathbf{r}_{2})}{r_{2}^{\prime 3}},
\end{equation}%
where $\mathbf{r}_{1}=$ and $\mathbf{r}_{2}$ are separation vectors between
point $\mathbf{r}$ of the surface and the particle and its image. The
boundary condition $\mathbf{n}_{t}=0$ gives two equations, i.e.,

\begin{equation}
(\mathbf{d}_{t}\cdot \mathbf{r}_{1})+(\mathbf{d}_{t}^{\prime }\cdot \mathbf{r%
}_{2})=0,\;t=x,y.  \label{dr}
\end{equation}%
We will consider the wall with the planar and homeotropic director alignment
individually.

\paragraph{Planar wall.}

The wall with a planar director alignment, Fig.4, coincides with the $yz$%
-plane $x=0$, while the $x$-axis is normal to the wall. Obviously, if $%
\mathbf{r}_{1}=(-x,y,z),$ then $\mathbf{r}_{2}=(x,y,z)$. The two equations (%
\ref{dr}) then are solved by $\mathbf{d}_{x}=\mathbf{d}_{x}^{^{\prime }}$
and $\mathbf{d}_{y}=-\mathbf{d}_{y}^{^{\prime }}$, Fig 4. The interaction
energy of the dipole dyad $(\mathbf{d}_{x},\mathbf{d}_{y})$ and its image $(%
\mathbf{d}_{x},-\mathbf{d}_{y})$ is readily calculated from eq.(\ref{Udd})
by substituting $\mathbf{d}_{x}=(d,0,0)$ and $\mathbf{d}_{y}=(0,d,0).$ The
force is obtained by differentiating this expression with respect to $%
\mathbf{R}$ at $R=2h.$ This gives a repulsive force with the magnitude

\begin{equation}
F_{d,planar}=4\pi K\frac{27d^{2}}{32h^{4}}.  \label{Fplan}
\end{equation}

\paragraph{Homeotropic wall}

In the homeotropic geometry, Fig. 5, the uniform director and the $z$-axis
with the onset at the wall are normal to the wall which coincides with the $%
xy$-plane. Obviously, if $\mathbf{r}_{1}=(x,y,-z)$ then $\mathbf{r}%
_{2}=(x,y,z)$. The two equations (\ref{dr}) then are solved by $\mathbf{d}%
_{x}=-\mathbf{d}_{x}^{^{\prime }}$ and $\mathbf{d}_{y}=-\mathbf{d}%
_{y}^{^{\prime }}$, Fig 5. The force is repulsive and has the magnitude

\begin{equation}
F_{d,hom}=4\pi K\frac{9d^{2}}{16h^{4}}.  \label{Fhom}
\end{equation}

This force is $1.5$ times weaker than $F_{d,planar}.$

\section{Conclusion}

The nematostatics in 2 and 3 d is very different. The former is very similar
to the 2d electrostatics where disclination cores are in place of electric
charges. The latter is similar to the electrostatics only in that its Green
functions are Coulomb-like. In 3d the counterpart of the electric charge
density is a dyad, the elastic charge can be induced only by an external
torque whose components play the role of an elastic charge dyad. In this 3d
colloidal nematostatics, the Coulomb-like interaction has the reverse sign.
We described some implications of the colloidal nematostatics in 3d and
showed that, in contrast to the electrostatics, the charges and dipoles are
repelled from the wall. One interesting effect is that, applying the
field-induced torque on a colloid, one can charge it, Fig.1. If the colloid
is an elastic dipole, then by applying the external field one can switch the
repulsion from the nematic surface from $1/h^{4}$ to $1/h^{2}$ regime. Our
results prompt the experimental tests of the interaction in nematic
emulsions that, rather than dealing with a pair of particles, can deal with
a single colloid at a wall.

\end{document}